\documentclass[letterpaper]{article}
\usepackage{amsfonts,amssymb,amsmath,bm,bbm,euscript,mathrsfs,calligra}
\usepackage[usenames]{color}
\usepackage{color,soul}
\usepackage[utf8]{inputenc}
\usepackage[T1]{fontenc}
\usepackage{textcomp}
\usepackage{graphicx}
\usepackage[tight,sf]{subfigure}
\usepackage{listings}
\usepackage{hyperref}
\usepackage{bookmark}
\usepackage{authblk}

\DeclareMathAlphabet{\mathcalligra}{T1}{calligra}{m}{n}
\DeclareFontShape{T1}{calligra}{m}{n}{<->s*[2.2]callig15}{}

\begin{document}
\sf

\begin{center}
   \vskip 2em  {\huge \sf  
Conformal Mechanics of Space Curves }\\
\vskip 3em
{\large \sf  Jemal Guven }

\vskip 2em

\em{Instituto de Ciencias Nucleares\\
Universidad Nacional Aut\'onoma de M\'exico\\
Apdo. Postal 70-543, 04510 M\'exico, DF, MEXICO}
\end{center}

 \vskip 1em
 
\begin{abstract}
\sf \small
Any conformally invariant energy associated with a curve possesses 
tension-free equilibrium states which are self-similar. When this energy is the three dimensional conformal arc-length, these states are the natural spatial generalizations of planar logarithmic spirals. In this paper, a geometric framework is developed to construct these states explicitly using the conservation laws associated with the symmetry. The tension along a curve, conserved in equilibrium, is  first constructed.  While the tension itself is not invariant, the statement of its conservation is. By projecting the conservation laws along the two orthogonal invariant normal directions, the Euler-Lagrange equations are reproduced in a manifestly conformally invariant form involving the 
conformal curvature and torsion. 
The conserved torque, as well as scaling and special conformal currents implied by the symmetry are constructed explicitly.  The special conformal current vanished with respect to an appropriate origin
in all tension-free states. 
 A sketch is provided of how self-similar spirals describing tension-free states
 can be constructed by integrating the conservation laws. The details will be provided 
in a companion paper. 
\end{abstract}

\vskip 1em


\vskip 3em

\section{Introduction}

Self-similar patterns often arise appears in physical processes, albeit only as an approximation or emergent within a larger process.  This may be a one-dimensional pattern within this larger system whose overall morphology is much more complicated. Thus the pattern of growth displayed by the chambers of a mollusk shell or the spiral arms in our galaxy are self-similar, while the detailed morphology itself is not. For this to happen, the effective geometrical degrees of freedom describing the self-similar spiral must somehow decouple from the rest of the  system. 
\\\\
In this paper, a simple variational scenario is presented in which self-similar curved patterns arise.
This involves associating a scale invariant energy with curves and examining its equilibria. Unless the tension vanishes,  these curves 
will not be self-similar. For if it does not vanish, a fundamental length scale is introduced, inconsistent with the self-similarity.  This suggests that the most appropriate construction involves a close inspection of the mechanical properties of self-similar equilibrium curves. 
\\\\
The first task is to identify an appropriate  energy consistent with this symmetry, ideally the simplest.  Even if this energy itself does not possess any obvious physical significance,  it will direct us towards the simplest Euler-Lagrange (EL) equations exhibiting the symmetry. \footnote{\sf
An analogous example for surfaces is provided by minimal surfaces with vanishing mean curvature.  While, of course,  they do occur as  area minimizing surfaces, they are often observed because they also minimize higher order functionals of the geometry,  which are difficult to access.} 
\\\\
There are, of course, any number of local scale invariant energies one can associate with a space curve 
using the undifferentiated Frenet curvature and torsion, $\kappa$ and $\tau$. They will be of the general form
\begin{equation}
H_0= \int ds \, \kappa \, \mathcal{G}(\kappa/\tau)\,,
\label{eq:H0def}
 \end{equation}
where $\mathcal{G}$ is any function, more familiar in a relativistic setting;
but such energies do not admit self-similar geometries, unless trivially. Indeed it is relatively straightforward to show that their critical points  are helices. One needs to raise the number of derivatives to identify an energy with non trivial self-similar equilibrium states.
 \\\\
If one admits first derivatives, the number of local scale invariants exhibiting non-trivial tension-free states increases dramatically.  There will be a conserved current associated with the scaling symmetry but  constructing these states is generally not a simple task.  
There is, however, one scale invariant energy that is very special. This is 
the conformal arc-length, given by  \cite{Sharpe1994,Musso1994}, 
\begin{equation}
H= \int ds \, \big( {\kappa'}^2 + \kappa^2 \tau^2  \big)^{1/4}\,.
\label{eq:Hdef}
 \end{equation}
Here $s$ is arc-length and prime denotes a derivative with respect to $s$. 
Significantly $H$ possesses logarithmic spirals as planar equilibrium curves \cite{Bolt}.
Their treatment as tension-free curves was developed in  reference \cite{Paper1}).
\\\\
$H$ is not the simplest energy one can write down at this order: 
the energy  $\int ds \, |\kappa'|^{1/2}$, the naive spatial generalization of its planar counterpart,
is simpler. The difference is that $H$ is  not just scale invariant, it is also conformally invariant, unchanged under transformations preserving angles,  
and it is the simplest possible conformal invariant of a curve.  Furthermore, at this order in derivatives,
it is also the unique such invariant.  The energy $\int ds \, |\kappa'|^{1/2}$ is not conformally invariant.
Notably, none of the energies of the form (\ref{eq:H0def}), except $\int ds \,\tau$, is conformally invariant (modulo $2\pi$); and no others, apart from $\int ds \,\kappa$, arise naturally in the context of curves. 
\\\\\
The role of conformal invariance was already appreciated in the context of biological growth by D'Arcy Thompson more than a hundred years ago and plays a significant in his groundbreaking work  \textit{On Growth and Form} \cite{Thompson}.  
Conformal transformations are the compositions of similarities (Euclidean motions and scaling) and inversion in spheres.
The additional symmetry associated with inversion  introduces an additional conserved \textit{special conformal} current.  
Just as the tension is the Noether current associated with translational invariance, this current is 
associated with invariance under translations in the inverted space. If the tension vanishes, this current will also. The consequences will be explored in this paper.
Neither the tension nor the special conformal current are conformal invariants.
Conformal symmetry will thus generally be broken in self-similar equilibrium states.  
\\\\
Conformal invariance is more familiar in the study of membranes.
The symmetric bending or Willmore energy of a   
two-dimensional  surface, quadratic in the extrinsic curvature \cite{Willmore}, first identified by 
Sophie Germain \cite{Sophie}, is a conformal invariant.  
The important role it plays in current membrane biophysics on the mesoscopic scales in which the membrane morphology comes into focus  cannot be overstated. 
 This was demonstrated spectacularly in a series of papers in the early nineties \cite{Seifert91,Kusner92,Julicher93}.  
 A  well-known review, capturing the heady advances at that time, is provided in reference  \cite{Seifert95}. For a recent review,  approaching the problem from a point of view not altogether different from the one used to approach the problem addressed here, see reference \cite{Udine2018}. In contrast, the analog of the bending energy along curves, the Euler Elastic energy, quadratic in the Frenet curvature, is not even scale invariant never mind conformally invariant (the following references provide points of entry into a now vast literature \cite{
LangerSinger,IveySinger,SingerSantiago,
Levien}).   
\\\\
Conformally equivalent curves are
characterized by their conformal torsion and curvature, second and third order respectively in derivatives of their Frenet  counterparts \cite{Sharpe1994,Musso1994}:
\begin{subequations} 
\label{eq:calKTdef}
\begin{eqnarray}
\mathcal{T}  &=& 
\frac{\kappa^3 \tau^2}{\nu^{5/2}}\, 
\left[\frac{\tau}{\kappa} - \left(\frac{\kappa'}{\kappa^2 \tau}\right)' \,
\right]\,;\\
\mathcal{K} &=&\frac{1}{ 8 \nu^3}\, \Big[4 \nu  (\nu''- \kappa^2 \nu)  - 5 {\nu'}^2 \Big] \,,
 \end{eqnarray}
 \end{subequations}
where $\nu= \big( {\kappa'}^2 + \kappa^2 \tau^2  \big)^{1/2}$.
The former vanishes if the curve is planar or spherical.
Logarithmic spirals and any curve connected to a logarithmic spiral by a conformal transformation are characterized by constant conformal curvature $\mathcal{K}$ (with $\tau$ set equal to zero in Eq.(\ref{eq:calKTdef}b)).  
\\\\
Log spirals are obviously not the only planar critical points of the  energy (\ref{eq:Hdef}). Their conformal descendants 
are also stationary, with the same energy; generically, these are double $\mathcal{S}$-shaped spirals which are self-similar. Thus
neither stationarity, nor constant conformal curvature selects the feature of logarithmic spirals that
sets them apart from their conformal descendants: what does is the fact that the conserved tension vanishes in logarithmic 
spirals whereas it does not in double spirals. In this paper we lay down a framework for constructing 
all tension-free equilibrium states of $H$. The details will be fleshed out in reference \cite{Paper3}).
These self-similar spirals form the natural spatial analogues of logarithmic spirals. As such they extend in a significant way the taxonomy introduced by D'Arcy Thompson.
They are classified by two independent parameters, their scaling rate $S$ and the magnitude of the toque $M$. 
Unlike logarithmic spirals, they will generally exhibit complex internal structure.
 \\\\
The Euler-Lagrange equations, describing conformal \textit{geodesics}, were first derived by Musso in 1994 \cite{Musso1994}.
He showed that these equations can be cast in terms of the conformal curvature and torsion: 
\begin{subequations} 
\label{eq:ELKT}
\begin{eqnarray}
{\mathcal{T}} ^{\bullet\bullet}-\mathcal{T}^3 -  2\mathcal{K} \mathcal{T}  &=& 0 \,;\\
                                          \left(\mathcal{K} +  3 \mathcal{T}^2  /2 \right)^\bullet &=& 0\,,
 \end{eqnarray}
 \end{subequations}
where the bullet represents a derivative with respect to conformal arc-length, $\bullet=(1/ \nu^{1/2}) d/ ds$. 
On a plane with $\mathcal{T}=0$, equilibrium states are described by curves with constant conformal curvature, described in reference  \cite{Sulanke1981}.
\\\\
More recently Magliaro et al. extended Musso's analysis to higher dimensions \cite{Magliaro2011}.  The conformal symmetry is realized as the Lorentz group acting on a three-dimensional invariant subspace of a five dimensional Minkowski space; the question is
approached by adapting the method of exterior differential systems  (see, for example, reference \cite{Bryant2003}).  This mathematics is beautiful---but \textit{caveat lector}---the demands placed on one's preparation are not insignificant. 
\\\\
Because conformally equivalent curves are completely characterized by their curvature and torsion, from a mathematical point of view Musso, has in principle identified solutions up to this equivalence.  However,  tension plays no role in Eqs.(\ref{eq:ELKT}), nor is it clear how to isolate tension-free equilibrium states if the problem is approached this way, a necessity if the equilibrium is to be self-similar. An approach that places tension at its center is needed.  
To accommodate the constraint on the tension, the conservation laws associated with conformal invariance are first constructed. One may then focus on how the vanishing tension propagates through these conservation laws.  Framed this way the geometric problem is transformed into a mechanical  one and the language of Euler elasticity, extended to accommodate the additional symmetry,  becomes appropriate.  The method of auxiliary variables developed by the author  will be used to identify these conservation laws \cite{auxil}.  This approach was used, in the simpler context of planar curves, in reference \cite{Paper1}. 
\\\\
The translational invariance of $H$ permits  
the EL equations to be recast as a conservation law: the Noether tension $\mathbf{F}$
is preserved along conformal \textit{geodesics}. This tension should \textit{not} be 
confused with the tension in an elastic rod.  Because this approach breaks manifest conformal invariance, it is not obvious that these equations are equivalent to Eqs.(\ref{eq:ELKT}).  It will be shown, however, that it is always possible to project this conservation law along two independent invariant normal directions  to reproduce Eqs.(\ref{eq:ELKT}).
Whereas tension is not itself conformally invariant, its conservation is. It is worth noting that the two invariant directions involved in this identification do not  coincide with the Frenet normal and binormal. Indeed, it quickly becomes evident that the Frenet frame transforms in an unexpectedly complicated way under conformal transformations.  
\\\\
Rotational invariance identifies the conserved torque, $\mathbf{M}$. In the study of Euler elastic curves, the two Casimir invariants of the Euclidean group, $\mathbf{F}^2$ and $\hat{\mathbf{F}}\cdot\mathbf{M}$
provide the constants of integration parametrizing solutions in terms of elliptic integrals (see, for example, \cite{SingerSantiago}, or \cite{CCG2002}). 
If the tension vanishes, as it does in self-similar equilibrium states, then so do both invariants.  But,  when the tension vanishes, the dimensionless conserved magnitude of the  torque is not only rotationally invariant, it also becomes translationally invariant: $\mathbf{M}\cdot\mathbf{M}$  is now a legitimate Euclidean invariant.
The torque will establish the spiral
axis in tension-free states. 
\\\\
The two additional conserved currents have no Euler elastic analogs. These are the scaling current
capturing the scale invariance of $H$ , and the special conformal current 
capturing its invariance under translations in the inverted space implied by the invariance under inversion in spheres.  
\\\\
In general, scale invariance completely fixes the tangential tension, which itself determines the full tension in equilibrium states.  If the tension vanishes, the conserved scalar current completely determines the Frenet torsion in terms of the Frenet curvature.  If this is inputted into the statement of torque conservation, 
a quadrature for the variable $\kappa'/\kappa^2$ is provided.  This completes the construction of the Frenet data, known to be sufficient to construct the equilibrium curve. 
It would, however, be a mistake to stop here:  the vectorial special conformal current, vanishing when the tension does,  
places a strong constraint on equilibrium self-similar geometries that was not anticipated.
By examining appropriate projections of this vanishing vector, the structure of self-similar spirals is revealed to be described by a repeating nutating unit, expanding monotonically as it precesses about the torque axis. The details  will be presented in reference \cite{Paper3}.

\section{ Conformal arc-length}

The Frenet description of a space curve in terms of its acceleration and torsion 
may be intuitive, but as the number of derivatives increases, it becomes increasingly difficult to identify geometrically significant invariants built out of them, never mind determining how they behave under deformation. 
The approach adopted in \cite{ACG2001} and then in \cite{CCG2002} 
works because the focus was on simple functionals of the curvature and the torsion which do not involve derivatives of $\kappa$ or $\tau.$ To treat a functional involving higher derivatives, such as the conformal arc-length defined by Eq. (\ref{eq:Hdef}) it is useful to introduce a one-dimensional covariant
derivative that is invariant with respect to rotation of the two normals, treating the curvatures along the normal directions democratically.  This approach was touched on in reference \cite{ACG2001}; its potential advantages were suggested but were not pursued.  
\\\\
Consider then an arc-length parametrized curve $s\to \mathbf{X}(s)$ in three-dimensional Euclidean space with the inner product between two vectors denoted by a centerdot separating them. 
\\\\
Let prime denote a derivative with respect to arc-length,  so that $\mathbf{t}=\mathbf{X}'$ is the unit  tangent vector to the curve. Let  
$\{\mathbf{t},\mathbf{n}^1,\mathbf{n}^{2}\}$ be an orthonormal frame defined along this curve, adapted to the tangential direction.  The projection of the acceleration $\mathbf{t}'$ onto each of the two normal vectors $\mathbf{n}^I$ defines a curvature, $K^I$, $I=1,2$.    As the curve is followed, 
these vectors will generally also rotate into $\mathbf{t}$ and among themselves. This behavior is summarized in the structure equations for the curve:
\footnote{\sf These equations are the direct analogues of the Gauss Weingarten equations for surfaces. The difference is that arc-length provides a privileged parametrization for curves. While this choice entails the surrender of manifest reparametrization invariance, it is straightforward to restore. The implications of this choice  will be addressed in the context of the calculus of variations in section \ref{calculus}.}
\begin{subequations}\label{eq:Structure}
\begin{eqnarray}
\mathbf{t}' &=& - K^I \mathbf{n}_I\,;\\
D\mathbf{n}^I &=&  K^I \mathbf{t}\,,\quad I=1,2\,.
\end{eqnarray}
\end{subequations}
The one-dimensional covariant derivative $D$ appearing in Eq.(\ref{eq:Structure}b) is defined by
$D^I{}_J = \partial_s \delta^{I}{}_J + \omega^{I}{}_{J}$, where
$\omega^{IJ}=\mathbf{n}^I\cdot{\mathbf{n}^J}'=-\omega^{JI}$ 
is a one-dimensional \textit{spin} connection.  Under a local rotation of the normals $D \mathbf{n}^I$,  like 
$\mathbf{n}^I$, transforms as a vector; ${\mathbf{n}^I}'$ does not. This framework can be extended in an obvious way to curves in higher dimensions.
\\\\
Let $\{\mathbf{t}, \mathbf{N}, \mathbf{B}\}$ denote 
the Frenet frame. The acceleration is directed along $\mathbf{N}$, so that $K^1=\kappa$, where $\kappa$ is the Frenet curvature;  the curvature $K^2$  vanishes: $K^2=0$.  The spin connection is now identified as the torsion: $\omega^{1}{}_2 = -\tau= - \omega^{2}{}_1$.  The well-known fundamental result is that, modulo Euclidean motions, the two independent scalars $\kappa$ and $\tau$ completely determine the curve \cite{doCarmo}.  If one's priority were to 
trace curves for art's sake one could stop here; however, one is left in the dark as to the internal structure of the curve. 
\\\\
The only space curves of constant $\kappa$ and $\tau$ are helices. The analogue of this result when $s$ is replaced by $\ln s$ is  discussed  in reference \cite{Monterde}.  In a sense, sketched in \cite{CCG2002},  any space curve is  approximated locally by a helix almost everywhere.
\\\\
Conformal 
invariants of curves must first of all be Euclidean invariants.
As such, they can be expressed in terms of the Frenet invariants and their derivatives.
But classifying them this way runs into difficulties as soon as higher derivatives are contemplated. For, whereas the curvature $\kappa$ is of second order in derivatives, the torsion $\tau$ is an order higher. A consequence is that it is not obvious what  the 
natural scalars are using only these scalars as building blocks. For example, the oft-misused expression,  $\kappa^2 + \tau^2$,  may appear reasonable  but it does not possess 
any geometrical or physical significance along curves that we know of. It does however appear to play a role, in an appropriate parametrization, as an approximation of the bending energy 
of developable strips when  $\tau/\kappa$ is small \cite{Wunder}.  
\\\\
This shortcoming never arises in the normal rotation covariant description introduced in Eq.(\ref{eq:Structure}b), the relevant reparametization invariant scalars are formed using $K^I$ and its covariant derivatives, $DK^I, D^2 K^I$ and so on,  as building blocks. The simplest scalar in this approach, $K^I K_I/2$, coincides with the 
Euler energy density. Its analogue constructed using first derivatives is
$DK^I DK_I/2$.  To express this scalar in terms of Frenet variables, using the definition of $D$, one finds
\begin{equation}
\label{eq:D1}
D K^1 = \kappa'\,;\quad
D K^2 = \kappa \tau\,.
\end{equation}
It then follows that  
\begin{equation}
\label{eq:DK2FS}
DK^I DK_I= {\kappa'}^2 + \kappa^2 {\tau}^2\,.
\end{equation}
This is none other than the sum of the Frenet scalars that appears in the conformal arc-length, Eq.(\ref{eq:Hdef}). This is not a coincidence.

\subsection{Inversion of curves in spheres}

We will now sketch a procedure to identify conformally invariant energies for curves.  This involves examining the behavior of curvatures and their covariant derivatives under conformal transformations.  
\\\\
Conformal transformations are the transformations of space that preserve angles. 
In all dimensions higher than two, any conformal transformation can be constructed by taking  compositions of inversions
in spheres, Euclidean motions, and scalings. The similarity transformations, Euclidean motions and scaling, act in an obvious way on Eq.(\ref{eq:Structure}). The behavior under  inversion in spheres is less obvious. But once it is understood how these equations transform behave under inversion in a sphere,  determining the behavior under a more general conformal transformation becomes straightforward. 
\\\\
Under inversion in a unit sphere centered at the origin,  the point $\mathbf{X}$ on the curve embedded in the Euclidean space maps to the point $\bar{\mathbf{X}} = \mathbf{X}/|\mathbf{X}|^2$.  Technically, the sphere should possess a radius
 to preserve dimensions. This will be not be written explicitly to avoid notational clutter.
\\\\
It is straightforward to describe how arc-length, the curvatures and their derivatives, defined in Eqs.(\ref{eq:Structure}a) and (\ref{eq:Structure}b), transform (see, for example, reference \cite{ForceDipoles} or \cite{Udine2018} where an analogous discussion is presented for surfaces).  
Under inversion in a unit sphere located at the origin,  the tangent and  the normal vector transform 
as follows:
$ \bar{\mathbf{t}} \to |\mathbf{X}|^2 \mathrm{R}_\mathbf{X} \, \mathbf{t}
$, whereas 
$\mathbf{n}^I\to -\mathrm{R}_\mathbf{X}\,\mathbf{n}^I$,
where
$\mathrm{R}_\mathbf{X}$ is the linear operator, 
\begin{equation}
\label{eq:Rdef}
\mathrm{R}_\mathbf{X} =
{\sf 1} - 2 \hat{ \mathbf{X}} \otimes \hat{\mathbf{X}}\,,
\end{equation}
representing a reflection in the  plane passing through the origin, orthogonal to $\hat{\mathbf{X}}$.
Here ${\sf 1}$
is the identity and $\hat{\mathbf{X}}=\mathbf{
X}/|\mathbf{X}|$. As a consequence of the behavior of the tangent vector, the arc-length 
transforms 
$ds \to d\bar s =ds/ |\mathbf{ X}|^2$.
\\\\
As for the curvatures and connection,  one finds that 
$K^I\to \bar K^I$ and $\omega^I{}_J\to \bar\omega^I{}_J$, where 
\begin{equation}
\label{eq:Komegainversion}
\bar K^I= -|\mathbf{ X}|^{2} \left( K^I - 2 \,(\mathbf{
X}\cdot \mathbf{ n}^I)  / |\mathbf{ X}|^2\right)\,,\quad
\bar \omega^{I}{}_{J}= - |\mathbf{ X}|^{2} \omega^{I}{}_{J}\,.
\end{equation}
The curvature transforms non-trivially.
The connection transforms by a simple weight; 
as a consequence, the covariant derivative does also: $D\to D_{\bar{s}} = -|\mathbf{ X}|^{2} D$.
A derivation of Eqs.(\ref{eq:Komegainversion}) is provided in Appendix A. 
As described in Appendix B,  these expressions are considerably simpler than their counterparts for the Frenet frame. 

\subsection{  Conformal arc-length is the simplest conformal invariant }

Using Eq.(\ref{eq:Komegainversion}), it is easy to see that 
\begin{equation}
D_{\bar s}\bar K^I 
= - |\mathbf{ X}|^2 D\, \Big(|\mathbf{ X}|^{2} ( K^I - 2 \,(\mathbf{
X}\cdot \mathbf{ n}^I)  / |\mathbf{ X}|^2)\Big)
= -|\mathbf{ X}|^4  D K^I\,.
\label{DKI}
\end{equation}
As a consequence, the scalar
\begin{equation}
\label{eq:U}
|DK|^2 = D K^I D K_I
\end{equation} 
is a \textit{primary field}, transforming with conformal weight $|\mathbf{ X}|^8$, so that
\begin{equation}
H = \int ds \, |DK|^{1/2}
\label{eq:HDK}
\end{equation}
is a conformal invariant of curves, regardless of the dimension.
 Using Eq.(\ref{eq:DK2FS}) this invariant is identified as the conformal arc-length,
defined with respect to the Frenet frame in Eq.(\ref{eq:Hdef}). 
\\\\
Curiously, the covariant expression (\ref{eq:HDK}) for the conformal arc-length, simple as it may be, 
does not show up in a search of the literature. Notice that whereas $DK^I$ transforms by a weight under conformal transformations, this is not true of 
either $\kappa'$ or $\kappa \tau$ separately. Confirming the conformal invariance of  (\ref{eq:Hdef}) using the Frenet frame is somewhat less than immediate.  Having said this,  it is only fair to direct the reader's attention to the insightful construction of this invariant presented in reference \cite{Fuster}.

\section{Conformal Curvature }

The transformation of the normal vector $D K^I$, given by Eq.(\ref{DKI})  
implies that $\nu=|DK|$ also satisfies 
\begin{equation}
\bar \nu = |\mathbf{X}|^4 \, \nu\,.
\end{equation}
Taking two further derivatives, we find
\begin{eqnarray}
\frac{d \bar \nu}{d\bar s}  &=& -\Big(|\mathbf{X}|^{6} \nu'
+ 4 |\mathbf{X}|^4 (\mathbf{X}\cdot\mathbf{t})\, \nu \Big)\,;\nonumber\\
\frac{d^2 \bar \nu}{d\bar s^2}   &=& 
|\mathbf{X}|^{8} \nu''  + 10 |\mathbf{X}|^6 (\mathbf{ X}\cdot\mathbf{ t})\, \nu' \nonumber\\
&& \quad+
4\, |\mathbf{ X}|^4 \Big[
4(\mathbf{ X}\cdot\mathbf{ t})^2 +  |\mathbf{ X}|^2  
[1- K_J (\mathbf{ X}\cdot\mathbf{n}^J)] \Big]\, \nu  \,.
\label{DNnu}
\end{eqnarray}
It is now simple to show that that the function $4 \nu  (\nu''- \kappa^2 \nu)  - 5 {\nu'}^2$ is a primary field, 
transforming with conformal weight $|\mathbf{ X}|^{12}$;  the conformal curvature, defined by Eq.(\ref{eq:calKTdef}b), 
is thus identified as a conformal scalar \cite{Sharpe1994}.\footnote{\sf 
The   $\nu\nu'$ terms generated under inversion of the scalar $\mathcal{B}=4 \nu\nu''  - 5 {\nu'}^2 $ cancel, leaving
 \begin{equation}
\bar{\mathcal{B}} =
|\mathbf{ X}|^{12} \mathcal{B} + 
16 |\mathbf{ X}|^8 \Big[-4(\mathbf{ X}\cdot\mathbf{ t})^2 +  |\mathbf{ X}|^2  
[1- K_J (\mathbf{ X}\cdot\mathbf{n}^J)] \Big]\, \nu^2  \,;
\end{equation}
However, the problematic second term is identical to a term originating in the transformation of  $4\kappa^2 \nu^2$.} 
$\mathcal{K}$ depends on $|DK|$ and its first three derivatives. 
A more useful expression for 
$\mathcal{K}$,
\begin{equation}
\label{calKTmu}
\mathcal{K}= -\mu (\partial_s^2+ \kappa^2/2) \mu  +  (\mu')^2/2\,,
\end{equation}
 is possible in terms of the variable 
\begin{equation}
\mu=|DK|^{-1/2}= ({\kappa'}^2+ \kappa^2 \tau^2)^{-1/4}\,:
 \label{eq:mudef}
 \end{equation} 
with respect to which the denominator in $\mathcal{K}$ gets suppressed. 
\\\\
Just as the normal vector $D K^I$ and the scalar $\nu$ transform under inversion with the same weight, one see that 
successive covariant derivatives are in correspondence with derivatives of $\nu$.
It is thus evident that 
\begin{equation}
\mathcal{K}_0=\frac{1}{ 8 |DK|^{3}}\, \Big[4 D K_I (D^2- \kappa^2) D K^I - 5 (D^2 K)^2\Big] 
\label{calK0def}
 \end{equation}
is also a conformal invariant.  While $\mathcal{K}$ and $\mathcal{K}_0$  coincide 
along planar curves with  $\nu=\kappa'$,
they differ in three or higher dimensions.  In section \ref{Torsion}, it will be shown that the difference  $\mathcal{K}-\mathcal{K}_0$ 
is a positive definite conformal invariant, quadratic in the conformal torsion. 

\section{Conformal Torsion}
\label{Torsion}

Let 
$\varepsilon_{I J}$ is the Levi-Civita tensor in the $2$ dimensional normal space, then 
one can form both a  pseudo-vector $\mathcal{I}_{I} = \varepsilon_{IJ} D K^J$ and a pseudo-scalar  
\begin{equation}
\label{eq:I2N3}
\mathcal{I} = \varepsilon_{IJ} D K^I D^2 K^J \,;
\end{equation}
the later is also a conformal pseudo-scalar with conformal weight $|\mathbf{X}|^{10}$. 
The symmetric product of first  derivatives appearing in the transformation of 
$\mathcal{I}  $ vanishes on contraction with the Levi-Civita tensor.\footnote{\sf
This construction is analogous to that of the 
Frenet torsions in \cite{ACG2001} using $DK^I$ and $D^2K^I$ instead of $K^I$ and $D K^I$.} 
It is also evident that one can construct higher-derivative analogues in higher-dimensional Euclidean spaces: in four Euclidean dimensions,  the analogue of $\mathcal{I}$ is
$\mathcal{I} = \varepsilon_{IJL} D K^I D^2 K^J D^3 K^L$, whereas 
$\varepsilon_{IJL} D K^J D^2 K^L$ forms a pseudo-vector. 
\\\\
One can now construct a conformal invariant using $\mathcal{I}$ and the invariant one-form, $\mu^{-1} ds$. 
At this point it is advantageous to introduce the invariant unit vector in the normal space, $\mathbf{U}= U^I \mathbf{n}_I$ where 
\begin{equation}
\label{UIdef}
U^I= D K^I/|DK|\,,
\end{equation}
satisfying $\bar U^I = U^I$.
By construction $U^I D U_I=0$, and $D U^I$ is also a primary field:
$\bar D_{\bar s} \bar U^I = |\mathbf{X}|^2 D U^I$.
\\\\
Define $D_\mu =\mu D$. Now, under conformal inversion, the vectors  $(D_\mu)^n U^I$, $n=1,2,3,\cdots$ 
form primary fields of weight one.
\begin{equation}
\label{DDUttranf}
D_{\bar \mu} \bar U^I =  D_\mu U^I\,,\quad 
{D_{\bar \mu}}^2 \bar U^I = {D_\mu}^2 U^I \,,\quad \cdots\,.
\end{equation}
The conformal torsion (a conformal invariant) is defined by 
\begin{equation}
\label{T1def}
\mathcal{T}   =  
\varepsilon_{IJ} \, U^I D_\mu U^J = \varepsilon_{IJ}  DK^I D^2 K^J/ |DK|^{5/2} = \mu^5  \mathcal{I}  \,,
\end{equation}
where $\mathcal{I}$ is given by Eq.(\ref{eq:I2N3}).
The conformally invariant total torsion is now defined by  
\begin{equation}J = \int ds \mu^{-1} \, \mathcal{T} = \int ds \, \varepsilon_{IJ} \, U^I D U^J\,.
\label{Jdef}
\end{equation}
In three-dimensions, $D_\mu U^I$ can be expressed in terms of $U^I$ and $\mathcal{T}$:
\begin{equation}
\label{DUepsUT}
D_\mu U_I= - \mathcal{T}  \,\varepsilon_{IJ} U^J \,;
\end{equation}
one can then expand $D_\mu^2 U^I$ in terms of the two orthogonal normal vectors $U^I$ and $D_\mu U^I$:
\begin{eqnarray}
\label{D2UUDU}
D_\mu^2\, U_I 
&=&  - \mathcal{T}  ^2 U^I + \frac{(\mathcal{T}  ^2)^\bullet }{2\mathcal{T}  ^2} \,D_\mu U_I \,.
\end{eqnarray}
Here the identity  
\begin{equation}
\label{DU2}
  (D_\mu U)^2 =
\mathcal{T}^2 \,,
\end{equation}
following from Eq.(\ref{DUepsUT}), has been used. The bullet represents the derivative with respect to conformal arc-length.
As a consequence of (\ref{D2UUDU}), the magnitude  of the conformal second derivative can also be cast in terms of $\mathcal{T}$ and its first conformal derivatives:  
\begin{equation}
(D_\mu^2 \,U)^2 = \mathcal{T}^4 + {\mathcal{T}^\bullet}^2 \,.\label{D2U2}
\end{equation}
It is now possible to answer the question posed earlier:
what is the relationship between 
$\mathcal{K}_0$, defined by Eq.(\ref{calK0def}),
and $\mathcal{K}$ defined by Eq.(\ref{eq:calKTdef}
b). 
Note that ($U^I= DK^I/\nu$) 
 \begin{eqnarray}
\mathcal{K}_0 &=&
\frac{1}{ 8 \nu^3}\, \Big[4 \nu U^I (D^2- \kappa^2) \nu U^I - 5 (D (\nu  U)^2\Big] \nonumber\\
&=& \mathcal{K} +
\frac{1}{ 8 \nu}\, \Big[4 U^ID^2 U^I  - 5 (DU)^2 \Big]= \mathcal{K} - 
\frac{9 \mathcal{T}  ^2}{ 8 }\,,
 \end{eqnarray}
where 
 Eqs.(\ref{D2UUDU}), (\ref{DU2}) as well as the unitarity of $U^I$ have been used.
\\\\
It is instructive to recast the conformal torsion in terms of the Frenet curvature and torsion and their derivatives. The efficient way to evaluate second (and higher) covariant derivatives in the Frenet gauge is to proceed iteratively. 
Thus, for $D^2 K^I$,  use is made of the identity
\begin{equation}
D^2 K^I= (D K^I)' +  \omega^I{}_J D K^J\,,
\end{equation}
where $DK^I$ are given by Eq.(\ref{eq:D1}).
Using the identities Eqs.(\ref{eq:D1}) for the lower derivatives, 
one immediately identifies 
\begin{equation}
\label{eq:D2}
D^2 K^1 =\kappa'' - \kappa \tau^2\,; \quad
D^2 K^2 =   (\kappa \tau)'   + \kappa' \tau\,.
\end{equation}
$\mathcal{I}$, defined by Eq,(\ref{eq:I2N3}), can now be factorized,
\begin{eqnarray}
\mathcal{I}   &=&  D K^1 D^2 K^2 - D K^2
 D^2 K^1 \nonumber\\
 &=& \kappa'  [\kappa \tau'   +2 \kappa' \tau] -
 \kappa \tau (\kappa'' - \kappa \tau^2)\nonumber\\
 &=& 
\kappa^3 \tau^2\,  \left[\frac{\tau}{\kappa} - \left(\frac{\kappa'}{\kappa^2 \tau}\right)' \,
\right]\,.
 \label{IN3}
 \end{eqnarray}
The integrated torsion (\ref{Jdef}) can now also be cast in terms of the Frenet variables,
\begin{equation}J= \int ds \, |DK|^{1/2}\, \mathcal{T}  = \int ds\,  \frac{\kappa^3 \tau^2}
{{\kappa'}^2+ \kappa^2 \tau^2}\, \left[\frac{\tau}{\kappa} - \left(\frac{\kappa'}{\kappa^2 \tau}\right)' \,
\right]\,.
\label{IN3collect}
\end{equation}
Significant properties of $\mathcal{T}$ and $J$ are collected in \ref{Jprop}.

\section{Critical points of curvature energies}
\label{calculus}

One is now in a position to examine the behavior of the 
conformal arc-length under deformations of the curve,
$\mathbf{X}\to \mathbf{X}+ \delta \mathbf{X}$. 
\\\\
In general, translational invariance of the energy identifies 
the EL derivative 
with respect to $\mathbf{X}$ with the divergence of a stress tensor, 
identified with the tension $\mathbf{F}$.   Conformal geodesics then satisfy $\mathbf{F}'=0$. 
Whereas the energies of two curves related by a conformal transformation coincide, the tensions within them generally differ.  Tension-free curves are necessarily in equilibrium.  In contrast to the planar reduction of this problem, however, it is not obvious if every equilibrium state is conformally equivalent to a tension-free state \cite{Paper1}.
\\\\
Consider, more generally, a functional defined on an arc-length  parametrized space curve, $H[ \mathbf{X}]$. This can always be cast in the 
form
\begin{equation}H[ \mathbf{X}]=
\int ds\, \mathcal{H}(K^I,D K^I)\,,
\end{equation}
where $D K^I = K^I{}' + \omega^I{}_J K^J$, and $\omega$ is the normal connection defined below 
Eq.(\ref{eq:Structure}b). The conformal arc-length depends only on $DK^I$. A dependence on $K^I$ will be admitted, not only because it involves no extra effort but because it facilitates comparisons with Euler-Elastica and,  more significantly,  it permits the implications of reducing the symmetry, replacing conformal invariance by 
scale invariance, to be explored. For example, the energy
\begin{equation}H[ \mathbf{X}]=
\int ds\, (DK^I DK_I + \alpha (K^I K_I)^2)^{1/4}
\end{equation}
is scale invariant for any choice of $\alpha$; but conformally invariant only if $\alpha=0$.
In any case, whenever the energy involves higher derivative energies, dismantling the  
covariant derivatives in favor of the  Frenet scalars and their derivatives is not an optimal strategy for turning  the variational crank.  Treating the normal directions democratically simplifies the implementation of the calculus of variations. 
\\\\
The method of auxiliary variables will be adopted to examine the behavior of $H$ under small deformations. This approach was developed originally  
to examine surfaces \cite{auxil} (see also \cite{Udine2018} for a recent review, tailored to the two-dimensional bending energy).  
While originally developed for energies quadratic in curvature, there is no obstacle to considering energies involving higher derivatives (see, for example, \cite{Guven2005} or \cite{Graham2017},  a factor of two and a sign error in \cite{Guven2005} were corrected discreetly  in the treatment provided in \cite{Graham2017}). The implications of curvature derivatives  on the boundary of contact in the adhesion 
of membranes were also explored in \cite{Deserno2007} using this approach.
\\\\
The idea is to treat $K^I$ and $\omega^{IJ}$ as independent variables in $H$.  To do this in a consistent way, it is necessary to introduce Lagrange multipliers to impose the structure equations connecting them to $\mathbf{X}$ as constraints. The functional dependence on $\mathbf{X}$ itself, as well as the intermediate variables $\mathbf{t}$ and  $\mathbf{n}^I$ appears only within the constraints.
\\\\
Thus one constructs the constrained functional
\begin{eqnarray}
H_C[ \mathbf{X},\mathbf{t}, \mathbf{n}^I, K^I,\omega^{IJ},\dots]  
&=&  H[K^I,\omega^{IJ}]
\nonumber\\
&&\quad + \int ds
\,\left[ \frac{1}{2} T (1 - \mathbf{t}\cdot \mathbf{t} )
- H_I (K^I- \mathbf{t}\cdot  D \mathbf{n}^I)
- \mathcal{S}_{IJ}\,(\omega^{IJ} - \mathbf{n}^I \cdot 
\mathbf{n}^J{}') \right]\nonumber\\
&&\quad\quad +  \int ds \left[ \frac{1}{2} \lambda_{IJ} (\mathbf{n}^I\cdot
\mathbf{n}^J - \delta^{IJ}) -  f_I (\mathbf{n}^I\cdot \mathbf{t})
+ \mathbf{F}\cdot (\mathbf{t}-
\mathbf{X}') \right]\,. \label{eq:aux}
\end{eqnarray}
The details of this approach will be developed in the context of surfaces in a general Riemannian background in reference \cite{Armas}. Antecedents can be found in a specific higher-dimensional (surface) context in \cite{CS};  a comprehensive alternative direct approach (without Lagrange multipliers) was developed for surfaces in a Riemannian background in \cite{Armas13}.
\\\\
The EL equations for $K^I$ and $\omega^{IJ}$ identify
$H_I$ and $\mathcal{S}_{IJ}$ as the Euler Lagrange derivatives of the 
unconstrained functional $H[K^I,\omega^{IJ}]$ with respect to 
$K^I$ and $\omega^{IJ}$ respectively:
\begin{equation}
H_I= \frac{\delta H}{\delta K^I}\,,\quad  \mathcal{S}_{IJ}  = \frac{\delta H}{\delta \omega^{IJ}}\,. 
\end{equation}
Explicit expressions for $H_I$ and $\mathcal{S}_{IJ}$ for energies involving 
$K^I$ and $D K^I$  are
\begin{subequations}
\label{eq:HIdefSIJdef}
\begin{eqnarray}
H_I &=&
\frac{\partial \mathcal{H}}{\partial K^I} -
D \, \left(\frac{\partial \mathcal{H}}{\partial D K^I}\right) \,;\\
S_{IJ} &=& 
\left(\frac{\partial \mathcal{H}}{\partial D K^{[I}}  \right)\,K_{J]} \,.
\end{eqnarray}
\end{subequations}
The square brackets appearing on the rhs of Eq.(\ref{eq:HIdefSIJdef}b)
indicates antisymmetrization.
\\\\
The EL equations for $\mathbf{t}$ and $\mathbf{n}^I$ identify the tension $\mathbf{F}$ along the curve to be  given by 
\begin{equation}
\label{eq:Fdef}
\mathbf{F}=\big ( T - H_I K^I\big) \, \mathbf{t} - (D H_I - 2 \mathcal{S}_{IJ}  K^J)\, \mathbf{n}^I \,.
\end{equation}
To see this, notice that the EL equation for  $\mathbf{t}$,  following from
Eq.(\ref{eq:aux}), reads 
\begin{eqnarray}
\label{eq:ELt}
\mathbf{F} &=& T \,\mathbf{t} - H_I D\mathbf{n}^I  +
 f_I \,\mathbf{n}^I\nonumber\\
&=& ( T - H_I K_I ) \,\mathbf{t}  + f_I \,\mathbf{n}^I\,,
\end{eqnarray}
where equations of structure (\ref{eq:Structure}b), themselves implicit in the  constraints, are used on the second line.
The  counterpart of Eq.(\ref{eq:ELt}) for  $\mathbf{n}$ reads 
\begin{equation}
\label{eq:ELn}
f_I \mathbf{t} = -D (H_I \mathbf{t}) + 2 \mathcal{S}_{IJ} D \mathbf{n}^J + D \mathcal{S}_{IJ} \mathbf{n}^J+  
\lambda_{IJ} \mathbf{n}^J \,,
\end{equation}
or, equivalently,
\begin{equation}
\label{eq:ELn1}
(f_I + D H_I - 2 \mathcal{S}_{IJ}  K^J) \mathbf{t} - (H_I K^J    + D \mathcal{S}_{IJ}  - 
\lambda_{IJ} )\mathbf{n}^J =0 \,.
\end{equation}
The vanishing of the tangential component of this homogeneous equation implies $f_I= - (D H_I - 2 \mathcal{S}_{IJ}  K^J)$, reproducing the normal projection in Eq.(\ref{eq:Fdef}).  The vanishing of the 
antisymmetrized normal component implies the addition kinematical constraint 
\begin{equation}
\label{eq:DSHK}
D \mathcal{S}_{IJ} + H_{[I} K_{J]} =0\,,
\end{equation}
which captures the normal rotational invariance of $H$. 
It is simple to confirm that  the identifications (\ref{eq:HIdefSIJdef}a) and (b) are consistent with Eq.(\ref{eq:DSHK}).  
\\\\
In this framework, the functions $\mathbf{X}$ appear only in the tangency constraint. Modulo boundary terms, examined in the next section, 
\begin{equation}
\label{eq:delHX}
\delta_\mathbf{X} H_C [ \mathbf{X},\mathbf{t}, \mathbf{n}^I, K^I,\omega^{IJ},\dots]  
= \int ds\, \mathbf{F}'\cdot \delta\mathbf{X} \,.\end{equation}
But  this
implies that the bulk variation of $H$ itself is given by  
\begin{equation}
\label{eq:delHX0}
\delta_\mathbf{X} H [ \mathbf{X}] 
= \int ds\, \mathbf{F}'\cdot \delta\mathbf{X} \,.\end{equation}
It is thus clear that, modulo appropriate boundary conditions, 
$H$ is stationary  with respect to variations of $\mathbf{X}$ when $\mathbf{F}$  is a constant vector along the curve:
\begin{equation}
\label{eq:Fprime0}
\mathbf{F}'=0\,.
\end{equation}
Eq.(\ref{eq:delHX0}) identifies $\mathbf{F}$ as the tension in the curve. 
To understand the conservation law (\ref{eq:Fprime0}),  it is useful to examine its 
normal and tangential projections separately.
Let $F_\|$ and $F_{\perp\,I}$ be the corresponding projections of $\mathbf{F}$ so that
\begin{equation}
\label{Fdecomp}
\mathbf{F}= F_\| \,\mathbf{t} + F_{\perp\,I} \,\mathbf{n}^I\,.
\end{equation}
Using the structure equations (\ref{eq:Structure}a) and (b), the projections of 
$\mathbf{F}'$ along $\mathbf{t}$ and $\mathbf{n}^I$ are then given respectively by
\begin{subequations}
\label{eq:FperpFpar}
\begin{eqnarray}
\mathcal{E}_\perp^I &=& \mathbf{n}^I\cdot \mathbf{F}'=
D F_\perp^I - K^I\, F_\| 
\,,\quad I=1,2\,;\\
\mathcal{E}_\| &=&\mathbf{t}\cdot \mathbf{F}'=F_\|' + F_{\perp \,I} K^I 
\,.
\end{eqnarray}
\end{subequations}
 Notice that, on any curve,  $\mathcal{S}_{IJ} K^I K^J=0$, so that $\mathcal{S}_{IJ}$ never appears on the lhs of Eq.(\ref{eq:FperpFpar}b).  In equilibrium, $\mathcal{E}_\|=0$ as well as $\mathcal{E}_\perp^I=0$, $I=1,2$. 
\\\\
In this framework, the one remaining unknown in the definition of $\mathbf{F}$ is $T,$ the multiplier imposing the unitarity of $\mathbf{t}$ or, equivalently, flagging $s$ as arc-length.
Had the curve been parametrized arbitrarily,  it would have been necessary to introduce a one-dimensional metric. In such an approach, $T$ is identified as the stress associated with this metric.  The identity $
\mathcal{E}_\|=0$, where $\mathcal{E}_\|$ is given by 
(\ref{eq:FperpFpar}b), is then tautological: a consequence of the manifest reparametrization invariance of $H$. This approach is not generally optional for surfaces \cite{auxil}. But here there is a privileged 
parametrization by arc-length. The price paid is the breaking of manifest reparametrization invariance. A consequence is that the tangential EL equation is no longer satisfied identically. Its new role is to recover the 
multiplier $T.$ Stationary states are then characterized by the two EL equations: $\mathcal{E}_I=0$, $I=1,2$. 
\\\\
To determine $T,$  use Eq.(\ref{eq:Fdef}) to 
recast Eq.(\ref{eq:FperpFpar}b) in the form
\begin{equation}
\label{eq:FparT}
\mathcal{E}_\|= (T  - 2 H _I  K_I)'  +  H_I D K^I\,.
\end{equation}
Now note, using  the definition of $H_I$ (\ref{eq:HIdefSIJdef}a), that
\begin{eqnarray}
H_I DK^I &=& \left[\left(\frac{\partial \mathcal{H}}{\partial K^I}\right) -
D \, \left(\frac{\partial \mathcal{H}}{\partial D K^I}\right) \right] \,DK^I \nonumber\\
&=&
\mathcal{H}' - \left(
\left( \frac{\partial \mathcal{H}}{\partial D K^I}\right) \, D K^I \right)'\,.
\end{eqnarray}
Using this identity in Eq.(\ref{eq:FparT}), and integrating $\mathcal{E}_\|=0$, 
 $T$ is determined modulo a constant of integration, $\sigma$:
\begin{equation}
T  = - \mathcal{H}  +  2 H _I  K_I  - \sigma +
\left( \frac{\partial \mathcal{H}}{\partial D K^I}\right) \, D K^I \,.
\label{eq:T}
\end{equation}
The constant $\sigma$ is associated with the global constraint on the total arc-length implicit in this approach. This constraint is relaxed by setting $\sigma=0$.
By identifying $T$,  the reparametrization invariance that was temporarily suspended in choosing $s$ to parametrize the curve, is re-established. 
One can now express the tangential component of the tension appearing in the decomposition (\ref{Fdecomp}) completely in terms of the $K^I$ and their covariant derivatives:
\begin{equation}
F_\|  = - \mathcal{H}  +   H _I  K_I  +
\left( \frac{\partial \mathcal{H}}{\partial D K^I}\right) \, D K^I \,.
\label{eq:Fpar}
\end{equation}
In the Frenet gauge, Eqs. (\ref{eq:FperpFpar}a) and (b)
can be used to express the conservation laws in the form:
\begin{subequations}
\label{eq:FperpparFrenet}
\begin{eqnarray}
 (F_\perp^1)' -\tau \,F_\perp^2    - \kappa F_\| &=& 0\,; \\  
 (F_\perp^2)' + \tau \, F_\perp^1 &=& 0\,;\\
F_\|'  + \kappa  F_\perp^1 &=& 0\,.
\end{eqnarray}
\end{subequations}
Eq.(\ref{eq:FperpparFrenet}c) 
determines $F^1_\perp$ completely in terms of $F_\|$; 
Eq.(\ref{eq:FperpparFrenet}a) then determines $F^2_\perp$ in terms of $F_\|$.
Finally, Eq.(\ref{eq:FperpparFrenet}b) provides a third order equation for $F_\|$.
It is clear from this decomposition that a
sufficient condition that $\mathbf{F}=0$ in equilibrium is that
the tangential component vanishes, or $F_\|=0$. 
While this approach is of interest in principle, it is not a very useful approach  to solving these equations in practice. 

\section{The tension for conformal arc length}

Let $\mathcal{H}= \mathcal{F} (\mathcal{U})$, where $\mathcal{U}=|DK|^2$.
Eqs.(\ref{eq:HIdefSIJdef}) then read
\begin{subequations}
\label{eq:HS}
\begin{eqnarray}
H_I&=& - 2 D (\mathcal{F}_\mathcal{U} D  K_I)\,;\nonumber\\
\mathcal{S}_{IJ} &=& 2\mathcal{F}_\mathcal{U} D K_{[I}\,\, K_{J]} \,,
\end{eqnarray}
\end{subequations}
where $\mathcal{F}_\mathcal{U}=\partial \mathcal{F}/\partial \mathcal{U}$. Thus 
$F_\|$, defined by Eq.(\ref{eq:Fpar}),
is given by
\begin{equation}
F_\|= -\mathcal{F} + 2\mathcal{F}_\mathcal{U}\, \mathcal{U} - 2 D( \mathcal{F}_\mathcal{U}\,  D K^I) K_I 
\,.\end{equation}
The normal stress, $F_{\perp\, I}$, appearing in 
Eq.(\ref{eq:Fdef}), is given by
\begin{equation}
F_{\perp\, I}  =
2 D^2 (\mathcal{F}_\mathcal{U} D  K_I) +2 \mathcal{F}_\mathcal{U}  \kappa^2 \mathrm{P}_K D K_{I} \,,  
\end{equation}
where $\mathrm{P}_K=\delta^{I}_{J}- \hat K^I \hat K_J$ is the projector onto normal vectors orthogonal to $K^I$.
\\\\
In particular, for conformal arc-length, $\mathcal{F}= |DK|^{1/2}$,so that 
\begin{eqnarray}
\label{Fconfarc}
 2 \, \mathbf{F} &=&
-\Big(\mu^3 \kappa\kappa' \Big)'\,\, \mathbf{t} + \Big(
 D^2 (\mu U^I) + \mu \kappa^2 \mathrm{P}_K U^I \Big)\,\mathbf{n}_I
\,,
\end{eqnarray}
where $\mu$ is defined in Eq.(\ref{eq:mudef}), and $U^I$ is the unit normal vector
(given by \ref{UIdef}) parallel to $D K^I$. The simple expression for the tangential projection of $\mathbf{F}$  will be understood to be a consequence of the scale invariance of the energy. 
The first Casimir invariant of the Euclidean group is given by $F^2=\mathbf{F}\cdot\mathbf{F}$;  $F$  is not however a conformal invariant. 

\section{Recovery of the conformally invariant EL equations}

Even though $\mathbf{F}$ itself is not  conformally invariant, the EL derivative $\mathbf{F}'$ is. This is not manifestly obvious in a construction focused on $\mathbf{F}$ so it is worth confirming.
\\\\
To reproduce the manifestly covariant EL equations (\ref{eq:ELKT}) from the conservation law, the first step is to replace covariant derivatives with respect to arc-length everywhere they appear in Eq.(\ref{Fconfarc}) by covariant derivatives with respect to 
conformal arc-length, defined above Eq.(\ref{DDUttranf}). We have for the normal tension given by Eq.(\ref{Fconfarc}):
\begin{eqnarray}
2  F_\perp^I &=&
\frac{1}{\mu} D_\mu \frac{1}{\mu} D_\mu (\mu U^I) + \mu \kappa^2 \mathrm{P}_K U^I \nonumber\\
&=&\frac{1}{\mu}   D_\mu^2   U^I +
\frac{1}{\mu} [(D_\mu^2 \ln \mu)  U^I + (D_\mu \ln \mu) \, D_\mu U^I] 
+ \mu \kappa^2 \mathrm{P}_K U^I\,.
\end{eqnarray}
Now define $F_U=2\mu U^I F_{\perp\,I}$;  evaluating the projection of $F_\perp^I$, one finds
\begin{eqnarray}
F_U &=& - \mathcal{T}^2  +
D_\mu^2 \ln \mu + ( \mu^2 \kappa^2  - (\mu  K\cdot U)^2)\,;\nonumber\\
&=& - \mathcal{T}^2  - \mathcal{K}
 +  (D_\mu \ln \mu)^2/2 +  \left( \frac{1}{2} \mu^2 \kappa^2 - (\mu K\cdot U)^2\right)\,,\label{IUdef}
\end{eqnarray}
where the identity (\ref{D2UUDU}) is used to introduce the conformal torsion $\mathcal{T}$ in the first line and the identity (\ref{calKTmu}) has been rewritten in the form,
 \begin{equation}
\mathcal{K}
=-D_\mu^2 \ln \mu - k^2 \mu^2/2 +  (D_\mu \ln \mu)^2/2\,,
\label{calKmu} 
\end{equation}
to introduce the conformal curvature $\mathcal{K}$ on the second.
\\\\
The normal projection of the tension orthogonal to $U^I$,
$F_{DU}=2\mu D U^I F_{\perp\,I}$, is
\begin{eqnarray}F_{DU}&=& \mathcal{T} \mathcal{T}^\bullet 
+ (D_\mu \ln \mu) \, \mathcal{T}^2 -  (\mu  K\cdot U)( \mu K\cdot D_\mu U )\,.
\label{IDUdef}
\end{eqnarray}
The next step is to  project the two normal EL derivatives of $H$ with respect to $\mathbf{X}$, given by Eqs.(\ref{eq:FperpFpar}a), along $U^I$ and $DU^I$. For  the projection of $\mathcal{E}^I$ along $U^I$, one has
\begin{eqnarray}
2\mu^2 U_I\mathcal{E}^I&=&2D_\mu  (\mu U_I F_\perp^I) - 2 (\mu D_\mu U_I +\mu^\bullet U_I)   F_\perp^I
- 2 \mu^2 (U\cdot K)  F_\|\nonumber\\
&=&  F_U^\bullet - D_\mu \ln \mu\,  F_U -  F_{DU} + (\mu U\cdot K)^\bullet (\mu U\cdot K)
\,,
\end{eqnarray}
where the identity $2\mu F_\|= -  (\mu^2 \kappa \kappa^\bullet)^\bullet= -  (\mu U\cdot K)^\bullet$ is used in the last term.
Using the expressions (\ref{IUdef}) and (\ref{IDUdef}) for $F_U$ and $F_{DU}$; collecting terms and reusing the identity, (\ref{calKmu}), this screed collapses 
into the simple form
\begin{eqnarray}
\mathcal{E}_{U} & :=& 2\mu^2 U_I\mathcal{E}^I\nonumber\\
&=&- \left(\frac{3}{2} \mathcal{T}^2  + \mathcal{K}\right)^\bullet  + L_1\,,
\end{eqnarray}
where
\begin{equation}
L_1=  \frac{1}{2}( \mu^2 \kappa^2)^\bullet -  (\mu K\cdot U) (\mu K\cdot U)^\bullet
-  \Big(
\mu^2 \kappa^2 - (\mu K\cdot U)^2\Big)\, D_\mu \ln \mu
 +  (\mu  K\cdot U)( \mu K\cdot D_\mu U ) \,.
 \end{equation}
It is straightforward, albeit tedious, to confirm that $L_1$ vanishes.
The projection of $\mathcal{E}^I$ along $DU^I$ can be written 
\begin{eqnarray}
\mathcal{E}_{DU}= 2\mu^2 DU_I\mathcal{E}^I&=&2D_\mu  (\mu DU_I F_\perp^I) - 2 (\mu D_\mu^2U_I +\mu^\bullet D U_I)   F_\perp^I
+  (\mu U\cdot K)^\bullet (\mu D U\cdot K)
\nonumber\\
&=&  F_{DU}^\bullet - \mu^\bullet F_{DU} - I_{D^2U}
+   (\mu U\cdot K)^\bullet(D U\cdot K) 
\nonumber\\
&=& 
F_{DU}^\bullet - \mu^\bullet F_{DU} +  \mathcal{T} ^2 F_U -  \frac{\mathcal{T}^\bullet}{\mathcal{T}}\,F_{DU}
+   (\mu U\cdot K)^\bullet(\mu D U\cdot K) \,,
\end{eqnarray}
where Eq.(\ref{D2UUDU}) has been used as well as the definition of $\mathcal{T}$ to express the projection of the normal tension along $D^2_\mu U^I$ in terms of $F_U$ and $F_{DU}$:
\begin{equation}
I_{D^2U}:=2\mu (D_\mu^2 U^I ) F_I= -  \mathcal{T} ^2 F_U  +  \frac{\mathcal{T}^\bullet}{\mathcal{T}}\,F_{DU}\,.
\end{equation}
Using once again the  expressions (\ref{IUdef}) and (\ref{IDUdef}) for 
$F_U$ and $F_{DU}$, the
manifestly conformally expression
\begin{equation}
\label{Conf2}\mathcal{E}_{DU} = - 2\mathcal{K} \mathcal{T}    + \mathcal{T}^{\bullet\bullet} -\mathcal{T}^3
\end{equation}
follows.
The projections of the EL derivatives along $U_I$ and $D_\mu U_I$, are manifestly conformally invariant. The corresponding  EL  equations, $\mathcal{E}_{U}=0$ and
$\mathcal{E}_{DU}=0$, are also and  they coincide with Eqs.(\ref{eq:ELKT}a) and (b).   The conservation of the tension (which is not itself rotationally invariant never mind conformally invariant) is thus equivalent to the conformally invariant EL equations, derived first by Musso \cite{Musso1994} using a very different approach, reflecting different objectives. Notably,  the conserved tension plays no role in \cite{Musso1994}.

\section{Boundary variations and conservation laws}
\label{boundary}

The next task is to construct the torque associated with rotational invariance as well as the scalar and vector currents associated with conformal invariance. 
All three conserved currents play a role in the construction of tension-free states. 
\\\\
First collect the boundary terms that have accumulated in the variation of 
$H_C$ defined by Eq.(\ref{eq:aux}).  One has  
\begin{equation}
\label{eq:delHJ0def}
\delta H_C[ \mathbf{X},\dots]  =   \int ds\, \mathbf{F}' \cdot \delta\mathbf{X} +  \int ds\, \mathcal{J}'\,,
 \end{equation}
 where
 \begin{eqnarray}
\mathcal{J}
&=&  (H_J \, \mathbf{t} +  \mathcal{S}_{IJ} \mathbf{n}^I) \,\cdot \delta \mathbf{n}^J
 -  \mathbf{F}\cdot \delta 
\mathbf{X} 
+ \left(\frac{\partial \mathcal{H}}{\partial D K^I}\right)  \, \delta K^I 
 \,.\label{eq:calJdef}
\end{eqnarray}
Here $H_I$ is defined by Eq.(\ref{eq:HIdefSIJdef}a) and $\mathcal{S}_{IJ}$ by Eq.(\ref{eq:HIdefSIJdef}b).
The first three terms contributing to $\mathcal{J}$ originate in the variations of $\mathbf{X}$ and $\mathbf{n}^I$ when derivatives are peeled off the variation and collected in a derivative.  For the familiar Euler Elastic energy or any energy involving $K^I$ alone, $\mathcal{S}_{IJ}=0$ and ${\partial \mathcal{H}}/{\partial D K^I}=0$ and the two surviving terms complete the specification of the boundary term. For conformal arc-length, neither of these terms vanish. At this order in derivatives, 
there is no boundary contribution associated with variations of the spin connection. This is easily understood: $\omega^{IJ}$ always appears in the combination $DK^I$, one derivative lower than  $K^I$. 
\\\\\
If the curve is in equilibrium, the first  term appearing in Eq.(\ref{eq:delHJ0def}) vanishes and only the boundary terms survive. 

\subsection{Rotational invariance and torque conservation}

For rotations, defined by the axial vector $\mathbf{b}$,
$\delta \mathbf{X}= \delta \mathbf{b}\times \mathbf{X}$ and $\delta \mathbf{n}^I= \delta \mathbf{b}\times \mathbf{n}^{I}$, we have
\begin{eqnarray}
\delta_\mathbf{b} H_C
&=&  \delta \mathbf{b}\cdot \int ds\, \left(\mathbf{M}' - \mathbf {X} \times \mathbf{F}' \right)\,, \label{eq:cb}
\end{eqnarray}
where the torque is defined by
\begin{equation}
\label{eq:Mdef}
\mathbf{M}= \mathbf {X} \times \mathbf{F} - H_1  \mathbf{n}_2  + H_2 \mathbf{n}_1
- 2\mathcal{S}_{12} \mathbf{t}\,.
\end{equation}
In equilibrium, with $\mathbf{F}'=0$, $\mathbf{M}$  is conserved,
$\mathbf{M}'=0$.  Notice that $\mathbf{M}$ possesses the same dimensions as $H$. As such it is dimensionless if $H$ is scale invariant.\footnote{To be technically correct, it possesses the dimensions of energy.}
\\\\
For the conformal arc-length, the EL derivatives with respect to $K^I$ and $\omega_{12}$ are given by 
\begin{equation}
\label{HIS12}
H_I=
 - \frac{1}{2}\, D ( \mu^3  D  K_I) \,;\quad
2\mathcal{S}_{12}
=
- \frac{1}{2}\, \mu^3  D K_{2}\,\, K_{1} \,,
\end{equation} 
where $\mu$ is defined by Eq.(\ref{eq:mudef}), 
and (from (\ref{Fconfarc}), we identify $2 F_\|=-\Big(\mu^3 \kappa\kappa' \Big)'$.
With respect to the Frenet frame they read
\begin{subequations}
\label{eq:H12S12}
\begin{eqnarray}
2H_1 
&=& 
- (\mu^3  \kappa')'+\mu^3 \kappa \tau^2 \\
2 H_2 
&=& 
- (\mu^3 \kappa \tau)' -  \mu^3 \kappa' \tau = - (\mu^3\kappa^2\tau)'/\kappa\\
2\mathcal{S}_{12} 
&=&  
- \mu^3 \kappa^2 \tau/2\,. 
\end{eqnarray}
\end{subequations}
The bending moment is given by the second Casimir invariant of the Euclidean group, 
$\mathcal{M} = \mathbf{M}\cdot\hat{\mathbf{F}}$.
$\mathcal{M}$, like $F$, is not a conformal invariant.  But both Euclidean Casimir invariants vanish in tension-free states, so the issue is moot. In such states, however, $\mathbf{M}$ is translationally invariant and $M^2=\mathbf{M}\cdot\mathbf{M}$ is a Euclidean invariant. 

\subsection{Conformal invariance and its manifestations}

The treatment of the calculus of variations, thus far, has not exploited the conformal invariance of the 
\textit{energy}, and the additional conserved currents implied by this invariance.  The task now is to examine it consequences and identify these currents.  Begin with scaling.

\subsubsection{Scaling}
\label{scaling}

Rescaling the geometry, $\delta_\lambda \mathbf{X} =\lambda \mathbf{X}$,  one has  $\delta_\lambda \mathbf{n^I} =0$,  whereas 
$\delta_\lambda K^I = - \lambda K^I$.
Substituting into Eq.(\ref{eq:calJdef}), one identifies $\mathcal{J}= \lambda S$, where the scaling current $S$ is given by
\begin{equation}
\label{eq:Scaledef}
 S=  -\mathbf{F}\cdot 
\mathbf{X} 
 + S_D\,,\end{equation}
and 
\begin{equation}
\label{eq:SDdef}
S_D := -{\partial \mathcal{H}}/{\partial D K^I} \,  K^I  \,. 
\end{equation}   
For the conformal arc-length, with $\mathcal{H}=|DK|^{1/2}$,
${\partial \mathcal{H}}/{\partial D K^I}= \mu^3 DK_I/2$, so that
\begin{equation}
\label{eq:SDcal}
S_D = -
\mu^3\kappa \kappa'/2\,,
\end{equation}
where $\mu$ is defined in Eq.(\ref{eq:mudef}).
It follows from the identity (\ref{eq:delHJ0def}) that the current $S$ satisfies
\begin{equation}
\label{eq:Sprime}
S' = -\mathbf{F}'\cdot \mathbf{X}\,,
\end{equation}
whenever the energy is scale invariant. 
$S$ is conserved when $\mathbf{F}$ is. Note that, like the torque $\mathbf{M}$, $S$ is dimensionless. 
\\\\
Eq.(\ref{eq:Sprime}) is equivalent to the identity 
\begin{equation}
\label{eq:Fparaderiv}
F_\| = - S_D' \,,
\end{equation}
so that $F_\|$ is expressible as a derivative whenever $H$ is scale invariant (whether in equilibrium or not). It is simple to show that the identity (\ref{eq:Fparaderiv}) is
equivalent to the Euler scaling equation (cf. \cite{Paper1}.
Eq.(\ref{eq:Fparaderiv}) explains the surprisingly simple form of $F_\|$ in Eq.(\ref{eq:Fpar}).
\\\\
In a tension-free state, $S=S_D$ is a constant independent of $\mathbf{X}$. This equation places a constraint on the 
torsion $\tau$ in terms of the dimensionless ratio $\Sigma =-\kappa'/\kappa^2$. 
\begin{equation}
\label{eq:GamZ}
\tau^2 /\kappa^2 = Z^2 (\gamma - Z)\,,
\end{equation}
where $Z = \Sigma^{2/3}>0$, and $\gamma = (2S)^{-4/3}$.
Thus $\tau$ vanishes whenever $Z=\gamma$. If $Z=\gamma$ everywhere, the spiral is logarithmic.
\\\\
Modulo this constraint, the conservation of torque implies that $\Sigma$ satisfies a quadrature, involving the addition parameter $M,$ the magnitude of the torque. In reference  \cite{Paper3}, we will demonstrate that 
\begin{equation}
\label{eq:Quadrature}
\frac{\gamma^2}{4}  {\dot Z}^2 +   Z^2 (\gamma -Z) \Big[\gamma^2 Z^2  - (m^2 +1) Z  + \gamma \Big]=0\,,
\end{equation}
where $m=M/S$ and the overdot represents  $d/d\Theta= \kappa^{-1} d/ ds$.  $Z$ or $\Sigma$ can in turn be integrated to identify the Frenet curvature $\kappa$. The value $Z=\gamma$ where torsion changes sign is accessible only in \textit{supercritical} spirals with $m^2>\gamma^3$ or $4MS> 1$. 
\\\\\
It is now possible to construct the self-similar spiral from its Frenet data \cite{doCarmo}. The axis of this spiral is defined by the direction of $\mathbf{M}$.  In a scale invariant theory missing the extra inversion symmetry, this is as much as one can do.  However, the additional symmetry under inversion implies a very strong constraint on the self-similar structure, that is not manifest in the Frenet data.
 
\subsubsection{Special conformal current}

\begin{figure}
 \begin{center}
\includegraphics[height=8cm]{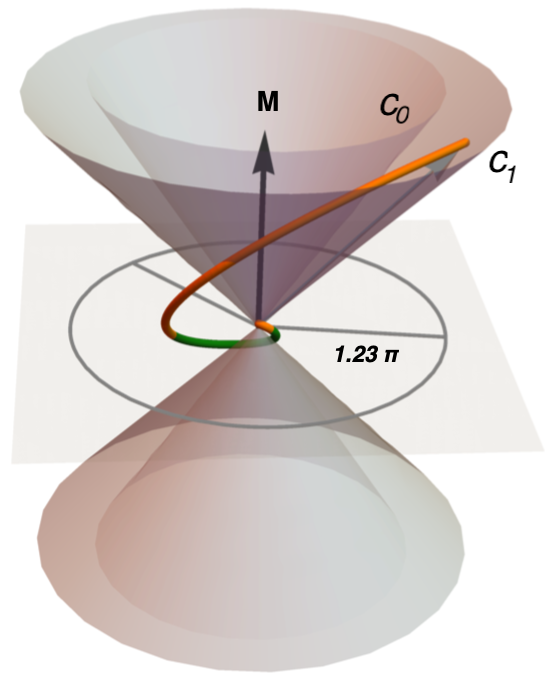}
\caption{\small \sf    
One complete cycle of a growing supercritical spiral $(\gamma=0.75,m=0.85)$), illustrating both the invariant cones ($\textit{C}_0$) restricting access to the poles,  as well as the pair of coaxial  outer cones ($\textit{C}_1$) along which the spiral reverses direction along the torque axis.  The segments lying above (below) the mid-plane orthogonal to $\mathbf{M}$ are colored orange (green). The spiral rotates by an angle $2.46\pi$ in one cycle. The angle rotated in the half-cycle between entry and exit of $\textit{C}_1$ (below the midplane and above it respectively) is indicated explicitly. This particular spiral dilates by a factor of $38.15$ over the course of this cycle.  Rotating this cycle by  $2.46\pi$, rescaling by a factor of $38.15$, and iterating generates the complete self-similar spiral.}
\label{Fig:Template}
\end{center}
\end{figure}

The identification of the second conformal current involves examining the behavior of a curve under a special conformal transformation, the composition of an inversion with a translation $\mathbf{X}\to \mathbf{X}+ \mathbf{c}$, followed
by a second inversion. Linearized in the intermediate translation, it is given by 
$\delta_\mathbf{c} \mathbf{X} = |\mathbf{X}|^ 2 \, \mathrm{R}_\mathbf{X} \,  \mathbf{c}$,
where $\mathrm{R}_\mathbf{X}$ is the linear operator on
$3$-dimensional Euclidean space defined along curves in Eq.(\ref{eq:Rdef}).
The vector $\mathbf{c}$ has dimensions
of inverse length squared.  
\\\\
In Appendix D,  the two identities 
\begin{equation}
\label{eq:dnc}
\delta_\mathbf{c} \mathbf{n}^I = 
2 (\mathbf{X}\cdot\mathbf{n}^I) \,\mathbf{c} - 2 (\mathbf{n}^I\cdot\mathbf{c})\, \mathbf{X}\,, 
\end{equation}
and
\begin{equation}
\label{eq:dKIc}
\delta_\mathbf{c}  K^I= 2 [ (\mathbf{X}\cdot\mathbf{c})\, K^I - (\mathbf{n}^I\cdot\mathbf{c})]
\end{equation}
are derived.
Using Eq.(\ref{eq:calJdef}), and the results just collected, the conformal current $\mathbf{G}$ associated with conformal arc length ($
\mathcal{J}=  \mathbf{G}\cdot \mathbf{c}$ in the identity (\ref{eq:delHJ0def})) is easily seen to be given by 
\begin{eqnarray}
\mathbf{G}&=& - 2 H_I \,  \mathbf{F}_0^I  + 2\mathcal{S}_{IJ} \, \mathbf{V}^{IJ} 
-   |\mathbf{X}|^2 \mathrm{R}_\mathbf{X} \mathbf{F} -  2 S_D\,  \mathbf{X} 
-\mu^3 DK_I \, \mathbf{n}^I\,.\label{eq:Gdef0}
\end{eqnarray}
Here $H_I$ and $S_{12}$ are given by Eq.(\ref{HIS12}) or, equivalently, in the Frenet variables by Eqs.(\ref{eq:H12S12}).  For reasons that will become apparent, the substitutions need not be made explicitly.
The identity ${\partial \mathcal{H}}/{\partial D K^I}= \mu^3 DK_I/2$
has been used as well as the
definition of $S_D$ given by  Eq.(\ref{eq:SDdef}); $ \mathbf{F}_0^I$ is defined by 
\begin{equation}
 \mathbf{F}_0^I =
(\mathbf{X}\cdot \mathbf{t})  \,\mathbf{n}^I - (\mathbf{X}\cdot\mathbf{n}^I) \, \mathbf{t} \,;
\end{equation}
the identity $\mathbf{n}^J\cdot \delta_\mathbf{c} \mathbf{n}^I = - \mathbf{c}\cdot \mathbf{V}^{IJ}$ has also been used, 
so that $\mathbf{V}^{IJ}$ is defined by
\begin{equation}
\label{eq:Vdef}
\mathbf{V}^{IJ} =    (\mathbf{n}^J\cdot\mathbf{X})\, \mathbf{n}^I
- (\mathbf{X}\cdot\mathbf{n}^I) \,\mathbf{n^J}  \,.
\end{equation}
Using Eq.(\ref{eq:Scaledef}) to reassemble the scaling current $S$, it is also possible to write
\begin{eqnarray}
\mathbf{G}&=& - 2 H_I \,  \mathbf{F}_0^I  + 2\mathcal{S}_{IJ} \, \mathbf{V}^{IJ} 
-   |\mathbf{X}|^2 \mathbf{F} -  2 S\,  \mathbf{X}
-\mu^3 \, D K_I \, \mathbf{n}^I
\,.\label{eq:Gdef}
\end{eqnarray}
One can now confirm that $\mathbf{G}$ satisfies 
\begin{equation}
 \mathbf{G}' = -   |\mathbf{X}|^2 \mathrm{R}_\mathbf{X} \mathbf{F}'=
 -2 S'\,  \mathbf{X} 
  -   |\mathbf{X}|^2 \mathbf{F}'\,,
  \label{eq:Gprime}
\end{equation}
consistent with Eq.(\ref{eq:delHJ0def}). 
$\mathbf{G}$ is conserved when $\mathbf{F}$ is.
The two identities, 
\begin{equation}
D \mathbf{F}_0^I = \mathbf{n}^I - \mathbf{V}^{IJ} \, K^J\,;\quad
D\mathbf{V}^{IJ} = - K^I \mathbf{F}_0^J + K^J \mathbf{F}_0^I\,.
\label{eq:DF0}
\end{equation}
simplify this demonstration.
Together, these identities can be viewed as the higher co-dimensional analog  for curves of an identity,  describing the normal vector in terms of a potential,  introduced in \cite{Laplace}.
\\\\
It is also possible to use the definition of the torque given by Eq.(\ref{eq:Mdef}),  
as well as the  three elementary identities 
\begin{equation}
\mathbf{F}_0^1=    (\mathbf{t}\cdot\mathbf{X})\, \mathbf{N}
- (\mathbf{N}\cdot\mathbf{X}) \,\mathbf{t} = \mathbf{X}\times \mathbf{B} \,;\quad
\mathbf{F}_0^2=    (\mathbf{t}\cdot\mathbf{X})\, \mathbf{B}
- (\mathbf{B}\cdot\mathbf{X}) \,\mathbf{t} = -  \mathbf{X}\times \mathbf{N}\,,
\end{equation}
and 
\begin{equation}
\mathbf{V}^{12} =    (\mathbf{B}\cdot\mathbf{X})\, \mathbf{N}
- (\mathbf{N}\cdot\mathbf{X}) \,\mathbf{B} = -   \mathbf{X}\times \mathbf{t} \,,
\end{equation}
in the Frenet frame, to express the first two terms on the right hand side of Eq.(\ref{eq:Gdef}) 
in terms of the moment of the excess torque, 
\begin{eqnarray}
 - 2 H_I \,  \mathbf{F}_0^I  + 2\mathcal{S}_{IJ} \, \mathbf{V}^{IJ} 
&=& 2 \mathbf{X}\times (\mathbf{M} - \mathbf{X}\times\mathbf{F}) 
\,.\label{eq:GdefM}
\end{eqnarray}
The conformal current  now assumes the form
\begin{eqnarray}
\mathbf{G}
= 2 \mathbf{X}\times (\mathbf{M} - \mathbf{X}\times\mathbf{F}) 
-   |\mathbf{X}|^2 \mathbf{F}-  2 S\,  \mathbf{X}
-\, \mu^3 (\kappa' \, \mathbf{N} +  \kappa \tau\, \mathbf{B}) 
\,,\label{eq:GdefMS}
\end{eqnarray}
involving the three conserved currents $\mathbf{F}$, $\mathbf{M}$, and $S$.
\\\\
Suppose that $\mathbf{F}=0$. Then, under translation $\mathbf{G}$ transforms by a constant vector.  It is thus possible to choose the origin so that $\mathbf{G}$, like $\mathbf{F}$, vanishes. This origin is the spiral apex.
But if $\mathbf{G}=0$, or
\begin{equation}
2 \mathbf{X}\times \mathbf{M}  
-  2 S\,  \mathbf{X}
-\, \mu^3 (\kappa' \, \mathbf{N} +  \kappa \tau\, \mathbf{B}) 
=0\,,\label{eq:GF0}
\end{equation}
all tension-free states are characterized completely by two independent parameters, the scaling rate $S$ and the torque magnitude,  $M.$ 
\\\\
The identity Eq.(\ref{eq:GF0}) has far-reaching implications for tension-free states. The details will be presented in reference \cite{Paper3}.
\\\\   
The internal structure of the spiral is captured by  
a single \textit{irreducible} repeating unit.  An example is illustrated in Figure \ref{Fig:Template}. 
In general, as shown in reference \cite{Paper3}, it is represented explicitly 
in the polar chart, $(\rho,\theta,\phi)$, adapted to the spiral apex 
and the direction of the torque by
\begin{equation}
\label{eq:kapparho}
(m^2  + 1)\kappa^2  \rho^2 =
\frac{\gamma}{Z }\, \left[ \gamma W_\pm(Z)^2 
+1 \right]\,,
\end{equation}
and
\begin{equation}
\label{eq:cosvsZ}
\cos\theta = {\sf sign}(\tau)\,
\frac{\sqrt{m^2 +1}}{m}\, \frac{
W_\pm(Z) }
{\sqrt{W_\pm(Z)^2
+ \gamma^{-1} }}
\,,
\end{equation}
where $Z$ is the solution of the quadrature (\ref{eq:Quadrature}), 
\begin{equation}
\label{eq:Wdef}
W_\pm ={\sf sign} ({\dot Z})  \, (Z-Z_-)^{1/2}(Z_+- Z)^{1/2} + Z^{1/2} 
(\gamma -Z)^{1/2}\,,
\end{equation}
and  
$Z_\pm$ are the two roots of the quadratic appearing in the quadrature 
(\ref{eq:Quadrature}). The azimuthal advance is given 
\begin{equation}
\phi = \frac{\gamma}{m}\, \int d\Theta \, 
        \frac{Z^{1/2}}{ 1 - \gamma m^{-2}  
W_\pm(Z)^2} \,.
\label{eq:phiTheta}
\end{equation}  
Here $Z$ is determined by the 
The value of $\kappa$, appearing in Eq.(\ref{eq:kapparho}), setting the \textit{scale} follows from the identity $-\dot \kappa/\kappa= Z^{3/2}$. 
\\\\
It can be shown that $\rho$ is always monotonic, whereas $\theta$ is bounded away from the torque axis by an
invariant cone ($\mathcal{C}_0$ in Figure 1).
The projection of $\mathbf{X}$ along the torque direction, $X_\|$,
is extremal when $Z=\gamma$,  which is where the torsion changes sign.
These extrema lie on a second invariant cone ($\mathcal{C}_1$ in Figure 1), coaxial with but outside $\mathcal{C}_0$.
The spiral nutates between consecutive extrema of $X_\|$,  each cycle of nutation corresponds to two periods in the potential well appearing in the quadrature (\ref{eq:Quadrature}).  
By scaling and rotating a  single cycle the complete self-similar spiral is generated.

\section{Conclusions}

A framework has been established for examining how self-similar space curves arise in the simplest scale invariant mechanics, described by the conformal arc-length.  The 
two conserved currents, the tension and the torque, associated with the underlying Euclidean invariance have been identified; the conservation of the tension
is equivalent to the EL equations. Even though the tension is not a conformal invariant, these equations can be cast in a manifestly conformally invariant form, involving 
the  two elementary conformal invariants of a space curve, the conformal curvature and torsion. 
Scale invariance was shown to constrain the tangential tension in terms of the scaling current. 
The additional conformal invariance implies the existence of a fourth current. The properties of this current 
have been emphasized. 
\\\\
A sketch, in broad outline, has been provided of the construction of a tension-free state. 
While it may not be obvious in what order the conservation laws should be integrated, a   
natural order suggests itself: the conserved scaling current determines the torsion in terms of the local curvature; the magnitude of the torque then provides a quadrature determining this curvature. 
When the tension vanishes, it was seen that 
the special conformal current does also if the origin is chosen appropriately. 
Not coincidentally this turns out to be the spiral apex. 
\\\\ 
The spiral will 
oscillate with increasing amplitude in successive cycles of nutation, twisting and precessing about the torque axis as it grows, the torsion changing sign where the growth along the axis is reversed. 
The nutating cycle captures the internal structure of the spiral. 
This structure possesses no analogue in logarithmic spirals where all points equivalent; individual points along the curve within spatial cycles are inequivalent. This internal structure may be evident within the Frenet data implied by the torque quadrature but its nature is not. The special conformal current plays a crucial role in teasing it out. 
\\\\
The mathematical significance of tension-free states stems from the fact that they represent the direct three-dimensional analogues of logarithmic spirals and arise as recurring motifs in self-similar processes.
In \cite{Paper3}, these states will be constructed in detail.  The compilation of a a taxonomy of optimal self-similar spatial spirals begins here. 
 
\section*{Acknowledgments}

Partial support from CONACyT grant no. 180901 is acknowledged. I have benefitted from discussions with Jay Armas on the details of the variational principle and with Denjoe O'Connor on conformal field theory. I also appreciate the hospitality of the Dutch Institute for Emergent Phenomena (DIEP) and the Dublin Institute for Advanced Studies during the summer of 2018 where elements of this work were presented as well as developed.  

\begin{appendix}

\setcounter{equation}{0}
\renewcommand{\thesection}{Appendix \Alph{section}}
\renewcommand{\thesubsection}{A. \arabic{subsection}}
\renewcommand{\theequation}{A.\arabic{equation}}

\section{Derivations of Eqs.(\ref{eq:Komegainversion})}

The derivation of Eq.(\ref{eq:Komegainversion}) involves the identity 
\begin{eqnarray}\mathrm{R}_\mathbf{X} \,{\mathrm{R}_\mathbf{X}}' &=&
-2 [{\sf 1} - 2 \hat{ \mathbf{X}} \otimes \hat{\mathbf{X}}]
[\hat{ \mathbf{X}} \otimes \hat{\mathbf{X}}' + \hat{ \mathbf{X}}' \otimes \hat{\mathbf{X}}]\nonumber\\
&=&
2 [ \hat{\mathbf{X}} \otimes \mathrm{P}_\mathbf{X} \mathbf{t}  - 
  \mathrm{P}_\mathbf{X} \mathbf{t}\otimes \hat{\mathbf{X}}]\,,\label{RRp}
\end{eqnarray}
where $\mathrm{P}_\mathbf{X}$ is the projection orthogonal to $\mathbf{X}$, which originates in the differentiation of the 
unit vectors appearing in $\mathrm{R}_\mathbf{X}$:
\begin{equation} \hat{\mathbf{X}}' = \mathrm{P}_\mathbf{X} \mathbf{t}/ |\mathbf{X}|\,.
\end{equation}
Using Eq.(\ref{RRp}) one has 
\begin{equation}
\label{RRpen}
\mathbf{t} \cdot \mathrm{R}_\mathbf{X}\,{\mathrm{R}_\mathbf{X}}' \,\mathbf{n}^I= 2( \mathbf{n}^I\cdot\mathbf{X})/|\mathbf{X}|^2\,,
\end{equation}
and also
\begin{equation}
\label{RRpnn}
\mathbf{n}^I\cdot \mathrm{R}_\mathbf{X}\,{\mathrm{R}_\mathbf{X}}' \, \mathbf{n}^J= 0\,.
\end{equation}
It is now clear that
\begin{equation}
\label{Kbarderiv}
\bar K^I= \bar{\mathbf{t}}\cdot \frac{ d \bar{\mathbf{n}}^I}{d \bar s}  =
- |\mathbf{ X}|^{2}  K^I + \mathbf{t}\cdot \mathrm{R}_\mathbf{X}{\mathrm{R}_\mathbf{X}}'  \mathbf{n}^I \,,
\end{equation}
which coincides, on using Eq.(\ref{RRpen}), with the expression for $\bar K^I$ given in  Eq.(\ref{eq:Komegainversion}). Similarly
\begin{equation}
\label{omegabarderiv}
\bar \omega^{IJ}= \bar{\mathbf{n}}^I\cdot\frac{ d \bar{\mathbf{n}}^I}{d \bar s}  =
- |\mathbf{ X}|^{2}  \omega^{IJ} + \mathbf{n}^I\cdot \mathrm{R}_\mathbf{X} {\mathrm{R}_\mathbf{X}}'  \mathbf{n}^J \,.
\end{equation}
The second term on the right vanishes because of the identity Eq.(\ref{RRpnn}). Eq.(\ref{omegabarderiv})
thus reproduces the expression for $\bar \omega^{IJ}$ given in  Eq.(\ref{eq:Komegainversion}). 

\setcounter{equation}{0}
\renewcommand{\thesection}{Appendix \Alph{section}}
\renewcommand{\thesubsection}{B. \arabic{subsection}}
\renewcommand{\theequation}{B.\arabic{equation}}

\section{Conformally transformed Frenet frame}

As a consequence of 
Eq.(\ref{eq:Komegainversion}), the Frenet curvature transforms as
\begin{eqnarray}
\label{bark}
\bar \kappa^2 &=&  \bar K_I \bar K^I \nonumber\\
&=& |\mathbf{X}|^2\kappa^2  - 2\kappa\, (\mathbf{N}\cdot\mathbf{X}) + |\mathbf{X}|^2 -(\mathbf{X}\cdot\mathbf{t})^2\,.
\end{eqnarray} 
It does not transform as
$\kappa\to  |\mathbf{X}|^2 \kappa  -   2(\mathbf{N}\cdot\mathbf{X})$, as might naively have been expected. Gauge choices do not transform in a simple way under conformal transformations.  The simplest way to determine how $\tau$ transforms
is to note that 
$U={\kappa'}^2 + \kappa^2 \tau^2 \to |\mathbf{X}|^8 U$.  Note in particular that  $\kappa'$ does not transform by a weight so neither does $\tau.$
It is evident that 
the Frenet frame transforms in a non-trivial way under conformal transformations, the more so  
the higher the dimension. 

\setcounter{equation}{0}
\renewcommand{\thesection}{Appendix \Alph{section}}
\renewcommand{\thesubsection}{C. \arabic{subsection}}
\renewcommand{\theequation}{C.\arabic{equation}}

\section{The conformal torsion}
\label{Jprop}

The term in square brackets appearing on the rhs of Eq.(\ref{IN3collect}) 
vanishes when
\begin{equation}
\frac{\kappa'}{\kappa^2 \tau} \left(\frac{\kappa'}{\kappa^2\tau} \right)' = \frac{\kappa'}{\kappa^2} \,,
\end{equation}
or
 \begin{equation}
\label{eq:taukapS2}
\tau= \frac{\kappa'}{\kappa (C\kappa^2 -1)^{1/2}}\,\,,
\end{equation}
where $C$ is a constant.
Curves on spheres are conformally equivalent to curves on planes, where $\tau=0$. Thus one should expect this invariant to vanish along such curves.
It is well-known that on a unit sphere,
the Frenet torsion is completely determined by the curvature and its first derivative through the relationship  (\ref{eq:taukapS2})  (see, for example, \cite{Gray}). On a unit sphere,
$C=1$.
\\\\
On a sphere, $\tau$ can equivalently be cast in terms of the geodesic curvature, $\tau= \kappa_g'/(\kappa_g^2 +1)$,  where we use the fact that the geodesic curvature $\kappa_g$ is related to $\kappa$  through the elementary relationship, $\kappa^2 = \kappa_g^2 +1$,
 between Frenet curvature, geodesic and normal curvatures applied to a sphere.\footnote{\sf This identity implies that the integrated torsion along a curve has the analytical expression in terms of the 
geodesic curvature,  $\int ds \, \tau= \arctan \kappa_g + c$.}
The rhs of Eq.(\ref{IN3collect}) thus vanishes if and only if  the curve lies on a sphere or a plane.  Notice that neither $\mathcal{K}$ nor $\mathcal{T}$  possess a definite sign. 
\\\\
An alternative reorganization of $\mathcal{T}$ points to a non-trivial result. 
Rewrite the middle line in (\ref{IN3}):
\begin{equation} \mathcal{I}   =
 \kappa'  [\kappa \tau'   + \kappa' \tau] -
 \kappa \tau \kappa'' + (\kappa'^2 + \kappa^2\tau^2) \tau\,. \end{equation}
Now $J$, given by  (\ref{IN3collect}),  can be decomposed in terms of the total torsion $\int ds\,\tau$ and an exact remainder:
\begin{eqnarray}
J  &=&  \int ds\,  \tau + \int ds \frac{\kappa^2 \tau^2}{{\kappa'}^2+ \kappa^2 \tau^2}\left(\frac{\kappa'}{\kappa \tau}\right)' =  \int ds\,  \tau + \int\, d\Phi \,,
  \end{eqnarray}
where $\tan\Phi= \kappa'/\kappa \tau$. 
This reproduces the well-known result that the total torsion is a conformal invariant mod $2\pi $  \cite{BanWhite75} (discussed in \cite{Sharpe1994}). This has interesting consequences in the context of self-similar spirals  \cite{Paper3}. The dimensionless variable $\Phi$ also plays an important role in the  construction of these spirals.

\setcounter{equation}{0}
\renewcommand{\thesection}{Appendix \Alph{section}}
\renewcommand{\thesubsection}{D. \arabic{subsection}}
\renewcommand{\theequation}{D.\arabic{equation}}

\section{Special Conformal Transformations}

{\large \bf Transformation of arc-length:}
\\\\
In general,
\begin{equation}
(|\mathbf{X}|^2 \mathrm{R}_\mathbf{X}\,\mathbf{c})'= 2 (\mathbf{X}\cdot\mathbf{t}) \,\mathbf{c} - 2 (\mathbf{X}\cdot\mathbf{c}) \,\mathbf{t}
- 2 (\mathbf{t}\cdot\mathbf{c})\, \mathbf{X}\,,
\end{equation}
so that
\begin{equation}
\label{eq:dsc}
\delta _\mathbf{c}  ds  =  -2\,  (\mathbf{X}\cdot\mathbf{c})\, ds \,.
\end{equation}
To see this,  describe arc-length along a curve parametrized by a fixed parameter $t$ so that $ds^2 = |\dot{\mathbf{X}}| dt^2$ and 
\begin{equation}
\delta ds^2  = 2 (\dot{\mathbf{X}}\cdot \delta \dot{\mathbf{X}}) dt^2 \,,
\end{equation}
where the dot represents differentiation with respect to $t$.
As a consequence,
\begin{equation}
\delta _\mathbf{c} ds^2 =2\,\mathbf{t} \cdot ( |\mathbf{X}|^2 \mathrm {R}_\mathbf{X} \mathbf {c})' \, ds^2 = - 4 (\mathbf{X} \cdot \mathbf{c} ) \, ds^2 \,.
\end{equation} 
\\\\
{\large \bf Transformation of basis vectors:}
\\\\
As a consequence of Eq.(\ref{eq:dsc}),
\begin{equation}
\label{dtc}
\delta_\mathbf{c} \mathbf{t} = 
2 (\mathbf{X}\cdot\mathbf{t}) \,\mathbf{c} - 2 (\mathbf{t}\cdot\mathbf{c})\, \mathbf{X}\,;
\end{equation}
Using $\mathbf{t}\cdot \delta_\mathbf{c} \mathbf{n}^I  = -\mathbf{n}^I\cdot \delta_\mathbf{c} \mathbf{t}$,  Eq.(\ref{eq:dnc}) 
follows.\\\\
{\large \bf Transformation of curvature, its derivative, and the normal connection:}
\\\\
In general, using the definition of $K^I$ and
Eq.(\ref{eq:Structure}a), one has 
\begin{equation}
\label{dKI}
\delta K^I= - \mathbf{n}^I\cdot 
\delta \mathbf{t}' + K_J \mathbf{n}^J\cdot \delta \mathbf{n}^I \,.
\end{equation}
Now use Eqs.(\ref{eq:Structure}a), (\ref{eq:dsc}), (\ref{dtc}) and (\ref{eq:dnc})  to express the special conformally transformed acceleration appearing in the first term,
\begin{equation}
\label{dtpc}
\delta_\mathbf{c} \mathbf{t}' = 
2 \mathbf{c} 
- 2 K^I (\mathbf{X}\cdot\mathbf{n}_I) \,\mathbf{c} + 2 K^I (\mathbf{n}_I \cdot\mathbf{c})\, \mathbf{X} -  2 (\mathbf{t}\cdot\mathbf{c})\, \mathbf{t}
-2 K^I (\mathbf{X}\cdot\mathbf{c})\,\mathbf{n}_I
\,.
\end{equation}
Using Eqs.(\ref{dtpc})  together with (\ref{eq:dnc}) in Eq.(\ref{dKI}) reproduces 
Eq.(\ref{eq:dKIc}).
\\\\
Note that the contribution proportional to $\mathbf{V}^{IJ} K_J$, defined by 
Eq.(\ref{eq:Vdef})  originating in the second term on the RHS of Eq.(\ref{dKI}) cancels and identical term 
appearing in the first.  
\\\\
In the same way,  it is found that
\begin{equation}\delta \omega^{IJ}= \mathbf{n}^I\cdot \delta \mathbf{n}^J{}' +
 \delta \mathbf{n}^I\cdot  \mathbf{n}^J{}' = 2 (\mathbf{X}\cdot\mathbf{c})\, \omega^{IJ}\,.
 \end{equation}
In conclusion\footnote{$
D\delta_\mathbf{c}  K^I= 2 (\mathbf{X}\cdot\mathbf{c})\, D K^I$.} 
\begin{equation}
\label{dDHIc}
\delta_\mathbf{c}  D K^I = 4 (\mathbf{X}\cdot\mathbf{c})\, D K^I \,.
\end{equation}
From Eq.(\ref{dDHIc}) and (\ref{eq:dsc}), follow the invariance of conformal arc-length, consistent  with the finite 
transformation  
(\ref{DKI})  under spherical inversion.\footnote{\sf 
$\delta_\mathbf{c}  DK_I D K^I =  8 (\mathbf{X}\cdot\mathbf{c})\, D K_ID K^I$, 
and 
$\delta_\mathbf{c}  \int ds \, (DK_I D K^I)^{1/4} = 0$.}

\end{appendix}

\end{document}